\def\year{2016}

\documentclass[letterpaper]{article}
\usepackage{aaai16}
\usepackage{array}
\usepackage{times}
\usepackage{helvet}
\usepackage{courier}
\usepackage{multirow}
\usepackage{balance}  
\usepackage{graphics}     
\usepackage{url}      
\usepackage{subfig}
\usepackage{rotating}

\usepackage[utf8]{inputenc} 

\begin{document}

\title{On the Behaviour of Deviant Communities in Online Social Networks}

\author{Mauro Coletto\\IMT Lucca\\CNR Pisa\\\textit{mauro.coletto@imtlucca.it} \And
 Luca Maria Aiello\\Yahoo\\\textit{alucca@yahoo-inc.com} \And
 Claudio Lucchese\\CNR Pisa\\\textit{claudio.lucchese@isti.cnr.it} \And
 Fabrizio Silvestri\\Yahoo\\\textit{silvestr@yahoo-inc.com} }

\maketitle

\begin{abstract}
\begin{quote}
On-line social networks are complex ensembles of inter-linked communities that interact on different topics. Some communities are characterized by what are usually referred to as \textit{deviant behaviors}, conducts that are commonly considered inappropriate with respect to the society's norms or moral standards. Eating disorders, drug use, and adult content consumption are just a few examples. We refer to such communities as \textit{deviant networks}. It is commonly believed that such \textit{deviant networks} are niche, isolated social groups, whose activity is well separated from the mainstream social-media life. According to this assumption, research studies have mostly considered them in isolation.
In this work we focused on adult content consumption networks, which are present in many on-line social media and in the Web in general. We found that few small and densely connected communities are responsible for most of the content production. Differently from previous work, we studied how such communities interact with the whole social network. We found that the produced content flows to the rest of the network mostly directly or through bridge-communities, reaching at least 450 times more users. We also show that a large fraction of the users can be inadvertently exposed to such content through indirect content resharing. We also discuss a demographic analysis of the producers and consumers networks. Finally, we show that it is easily possible to identify a few core users to radically uproot the diffusion process. We aim at setting the basis to study {\em deviant communities} in context.
\end{quote}
\end{abstract}

\section{Introduction} \label{sec:introduction}

The structure of a social network is fundamentally related to the interests of its members. People assort spontaneously based on the topics that are relevant to them, forming social groups that revolve around different subjects. This tendency has been observed with quantitative studies in several online social media~\cite{leskovec08planetary,aiello12friendship}. In the past, researchers have explored the relationship between information diffusion and network structure~\cite{barbieri13cascade}, focusing on the structural and dynamical properties of specific topical communities such as groups supporting political parties~\cite{conover11political}, or discussion groups about rumors, hoaxes~\cite{ratkiewicz10detecting} and conspiracy theories~\cite{bessi15science}.

Online social media are also favorable ecosystems for the formation of topical communities centered on matters that are not commonly taken up by the general public because of the embarrassment, discomfort, or shock they may cause. Those are communities that depict or discuss what are usually referred to as \textit{deviant behaviors}~\cite{clinard15sociology}, conducts that are commonly considered inappropriate because they are somehow violative of society's norms or moral standards. Pornography consumption, drug use, excessive drinking, eating disorders, or any self-harming or addictive practice are all examples of deviant behaviors. Many of them are represented, to different extents, on social media~\cite{haas10online,morgan10image,dechoundhury15anorexia}. However, since all these topics touch upon different societal taboos, the common-sense assumption is that they are embodied either in niche, isolated social groups or in communities that might be quite numerous but whose activity runs separately from the mainstream social media life. In line with this belief, research has mostly considered those groups in isolation, focusing predominantly on the patterns of communications among community members~\cite{gareth15people} or, from a sociological perspective, on the motivations to that make people join such groups~\cite{attwood05people}.

In reality, people who are involved in deviant practices are not segregated outcasts, but are part of the fabric of the global society. As such, they can be members of multiple communities and interact with very diverse sets of people, possibly exposing their deviant behavior to the public. In this work we aim to go beyond previous studies that looked at deviant groups in isolation by observing them \textit{in context}. In particular, we want to shed light on three matters that are relevant to both network science and social sciences: \textit{i)} how much deviant groups are structurally secluded from the rest of the social network, and what are the characteristics of their sub-groups who build ties with the external world; \textit{ii)} the extent to which content produced by a deviant community spreads and is accessed (voluntarily or inadvertently) by people outside its boundaries; and \textit{iii)} what is the demographic composition of producers and consumers of deviant content and what is the potential risk that young boys and girls are exposed to it.

In this initial study we undertake to answer those questions focusing on the behavior of \textit{adult content} consumption. Public depiction of pornographic material is considered inappropriate in most cultures, yet the number of consumers is strikingly high~\cite{sabina08nature}. Despite that, we are not aware of any study about the interface between adult content communities and the rest of the social network. We study this phenomenon on a large dataset from Tumblr, considering big samples of the follow and reblog networks for a total of more than 130 million nodes and almost 7 billion directed dyadic interactions. To spot the community that generated adult content, we also recur to a large sample of 146 million queries from a 7-month query log from a very popular search engine (Section~\ref{sec:methodology}), out of which we build an extensive dictionary of terms related to adult content that we make publicly available.

Results show that:
\begin{itemize}
	\item The deviant network is a tightly connected community structured in subgroups, but it is linked with the rest of the network with a very high number of ties (Section~\ref{sec:results:community}).
	\item The vastest amount of information originating in the deviant network is produced from a very small core of nodes but spreads widely across the whole social graph, potentially reaching a large audience of people who might see that type of content unwillingly. Although the consumption of deviant content remains a minority behavior, the average local perception of users is that neighboring nodes reblog more deviant content than they do (Section~\ref{sec:results:diffusion}).
	\item There are clear differences in the age and gender distributions between producers and consumers of adult content. The differences we found are compatible with previous literature on adult material consumption: producers are older and more predominantly male and age greatly affects the consumption habit, strengthening it in males and weakening it in females (Section~\ref{sec:results:demographics}).
\end{itemize}

 \section{Related Work} \label{sec:related}

\noindent \textbf{Groups in online social media.} 

\noindent Computer science research has dealt extensively with the problem of classification of groups along structural, temporal, behavioral, and topical dimensions~\cite{negoescu08Analyzing,grabowicz13distinguishing,aiello15group}. The relationship between group connectivity and shape of information cascades has also been explored, revealing an intertwinement between community boundaries and cascade reach that is particularly tight in communities built upon a common theme shared by all of their members~\cite{easley10networks,romero11interplay,barbieri13cascade,borregon14characterization}. The degree of inter-community interaction has been analyzed mostly in the context of heavily polarized networks, the most classical example being online discussions between two opposing political views~\cite{adamic2005political,conover11political,feller11divided}. These studies explored methods to quantify segregation~\cite{guerra13measure}, but mainly focus on networks formed by two main divergent clusters.

\hbox{}
\noindent \textbf{Deviant communities.} 

\noindent Deviant networks have been analyzed mostly in isolation. Studies about the depiction of drug and alcohol use in social media adopted mainly the content perspective. Researchers aimed at identifying the elements that boost content popularity, investigated the effect of gender on engagement, and studied the perceptions that deviant content arises in the young public~\cite{morgan10image}. Research has been conducted around anorexia-centered online communities~\cite{gavin08presentation,ramos11anorexia,boero12anorexia}, also on Tumblr~\cite{dechoundhury15anorexia}, investigating a wide range of aspects including the construction and management of member identities, the processes of social recognition, the emergence of group norms, and the use of linguistic style markers. Similar studies have been published over the years on communities of self-injurers and negative-enabling support groups, in which members encourage negative or harmful behaviors~\cite{haas10online}. Fewer studies touch upon network-related aspects. One notable example is the work by Gareth et al.~(\citeyear{gareth15people}) that provides an overview of behavioral aspects of users in the PornHub social network, with particular focus on the role of sexuality and gender. More loosely related are studies on the so-called \textit{dark networks}, mostly motivated by the need of finding effective methods to disrupt criminal or terroristic organizations~\cite{xu08topology}. The study by Christakis et al.~(\citeyear{christakis08collective}) about the communication network between smokers and non-smokers is one of the few quantitative studies that addresses the interaction between the social network and one of its sub-groups, but it strongly focuses on the phenomenon of contagion.

\hbox{}
\noindent \textbf{Adult content consumption.} 

\noindent In the context of internet pornography consumption, computer science literature studied the categorization of content and frequency of use~\cite{schuhmacher13exploring,tyson13demystifying,hald15types}. A wider corpus of research has been produced by social and behavioral scientists by means of surveys administered to relatively small groups. Special attention has been given to the relationship between age or gender and the exposure (voluntary or unwanted) to internet porn~\cite{sabina08nature,ybarra05exposure,buzzell05demographic,mitchell03exposure,chen13exposure}, with particular interest to the age band of young teens~\cite{mitchell03exposure,chen13exposure,wolak07unwanted}. Numbers vary substantially between studies, but clearly men are more exposed than women (approximately 75\%-95\% vs. 30\%-60\%), with men exposed more frequently~\cite{hald06gender} and women more often involuntarily. It is estimated that young teens that are often exposed accidentally (roughly 25\% to 66\% of the times) and are also exposed to violent or degrading pornography (20\% among female, 60\% among male)~\cite{romito15factors}. Researchers have also pointed out the potential harm that adult material consumption through internet can cause, including addiction~\cite{kuhn14brain} and increased chance of adopting aggressive behavior~\cite{allen95meta}. Exposition also correlates with drug use~\cite{ybarra05exposure} and with lack of egalitarian attitude towards the other sex~\cite{hald13pornography}. Although delving into the potential harm of pornography is far beyond the scope of our work, this inherent risks provide an additional motivation to focus on this particular type of deviant community.

 \section{Deviant graph extraction} \label{sec:methodology}

This study uses data collected from Tumblr, a popular micro-blogging platform and social networking website. The dynamics of the Tumblr community are based mostly on three possible actions. Users can \textit{post} new entries on their blogs usually containing multimedia content, \textit{repost} on their blogs any post previously published by others (similarly to Twitter retweets), and \textit{follow} other users to receive updates from their blogs in a stream-like fashion. Users might own multiple blogs, but for the purpose of this study we consider blogs as users, and we will use the two terms interchangeably.

We consider as {\em deviant nodes} those users who post content about a given {\em deviant topic}. To identify deviant nodes we resort to data from search logs. As shown in other studies~\cite{lee11cyberporn}, if a deviant query {\em hits} (i.e., leads to the click of) a Tumblr blog URL, then the blog is a candidate {\em deviant node}.

\begin{figure}[tp]
\centering
\includegraphics[clip=true, width=\columnwidth]{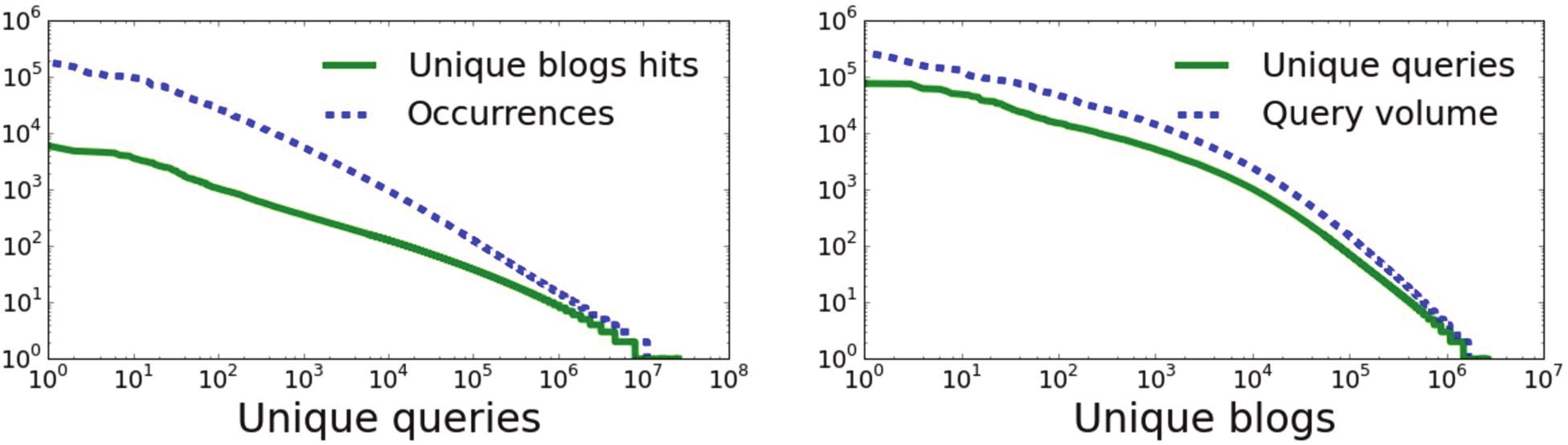}
\caption{Distributions of: (left) number of blogs hit by a query and number of occurrences of a query; (right) volume of (unique) queries hitting a blog.}
\label{fig:query_distributions}
\end{figure}

In our analysis we use a seven-month long query log (from Jan. to Jul. 2015) of a major search engine, from which we collected a random sample of 146M query log entries whose clicked URL belongs to the \url{tumblr.com} domain. We limit our study to queries that were submitted from the United States. After a simple query normalization process involving lowercasing and the removal of numbers, additional spaces, and of the word ``tumblr'' with its most common misspellings (as observed from the term distribution) we obtained about 26M unique queries that hit a total of 2.7M unique Tumblr blogs. As expected, the distribution of number of queries hitting a blog is very skewed, with most popular blogs being reached by hundreds of thousands of clicks originating from search queries (Figure~\ref{fig:query_distributions}). In the remainder of this work, we focus on adult content, this being a very common {\em deviant} topic on the Web. The same kind of analysis could be conducted on any other {\em deviant} topic.

To maximize the accuracy and coverage of the set of discovered deviant nodes, we devise an iterative semi-supervised  {\em Deviant Graph Extraction} procedure. Given a query log ${\cal Q}$, and a set ${\cal K}_i$ of deviant keywords (possibly multi-grams), we define as ${\cal Q}({\cal K}_i)$ the set of queries in ${\cal Q}$ that exactly match any of the keywords in ${\cal K}_i$. Based on the query log information, the set ${\cal Q}({\cal K}_i)$ yields a collection of clicked URLs from which we selected those corresponding to blogs in the Tumblr domain. We denote such set of blogs as ${\cal B}({\cal K}_i)$. To reduce data sparsity, we filter out the blogs  in ${\cal B}({\cal K}_i)$ with less than two unique incoming queries in ${\cal Q}({\cal K}_i)$ or less than 3 clicks originated by them.

The set of queries hitting ${\cal B}({\cal K}_i)$ is used to create a new set of keywords ${\cal K}_{i+1}$ and to re-iterate the procedure.  Given the current set of deviant nodes ${\cal B}({\cal K}_i)$ we identify the 10\% of blogs with highest proportion of query hits that match words in ${\cal K}_i$; those are the blogs that are hit mostly by deviant queries compared to other query types. We select all the unique queries that hit those blogs and merge them with ${\cal K}_i$, thus obtaining a new set of keywords ${\cal K}_{i+1}$, which is used to feed the next iteration of the algorithm. The procedure is repeated until the sizes of both ${\cal K}_i$ and ${\cal B}({\cal K}_i)$ converge.


The initial set ${\cal K}_0$ is obtained as follows. We first create a keyword set as the union of the search keywords from professional adult websites along with the list of adult performers published by movie production companies. To extend the coverage also to blogs that are reached predominantly by Spanish queries (the second most used language in US), we also translated to Spanish the initial set of keywords. From this initial set we manually extracted two dictionaries of respectively $5,\!152$ and $5,\!283$ search keywords (mono-grams, bi-grams, multi-grams), which were used to filter queries in the query log following two strategies: 1) \textit{exact match}, selecting those queries in the query log which match exactly one search keywords in the first dictionary, 2) \textit{containment}, selecting those queries subsuming any search keywords term in the second dictionary. For instance, the word \textit{porn} is not included in the containment dictionary because queries like \textit{food porn} should not to be detected as adult. The union of the queries detected by the two strategies hits a set of blogs, whose most frequent incoming queries were manually inspected to detect further $351$ search keywords. The union of these terms with the {\em exact match} dictionary leads to a set of $5,\!503$ {\em deviant queries} ($5,\!152$ + $351$) which is used as the seed set ${\cal K}_0$ to bootstrap the {\em deviant graph} extraction.

The above algorithm is biased towards the query log data, and on the popularity of blogs measured through the volume of search queries. On the other hand, this method allows to identify very quickly nodes that are likely to be relevant in the network as they produce the most interesting content to Web users. Also, as the procedure is network-oblivious (the graph structure is not exploited), no bias is introduced in our analysis of the network.

\begin{figure}[t]
\centering
\includegraphics[clip=true, width=.99\columnwidth]{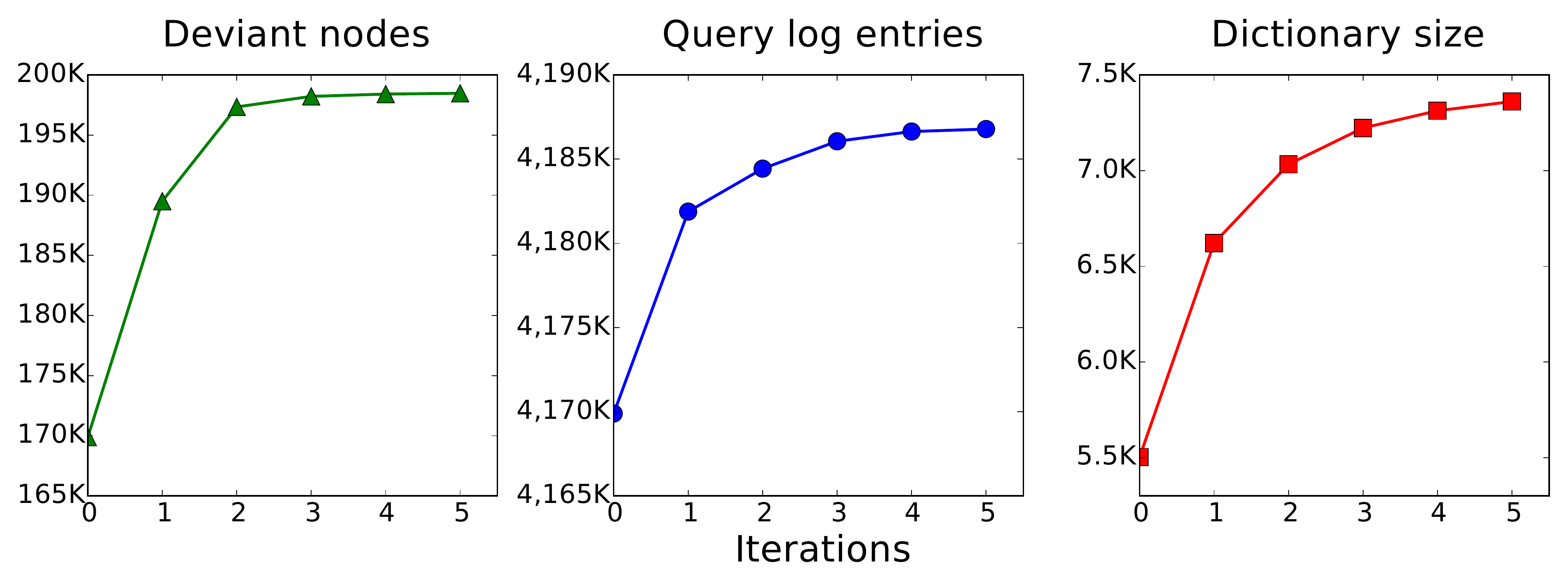}
\caption{Convergence of the three quantities used in the Deviant Graph Extraction procedure.}
\label{fig:query_expansion_iterations}
\end{figure}

\begin{figure}[t]
\centering
\includegraphics[clip=true, width=.99\columnwidth]{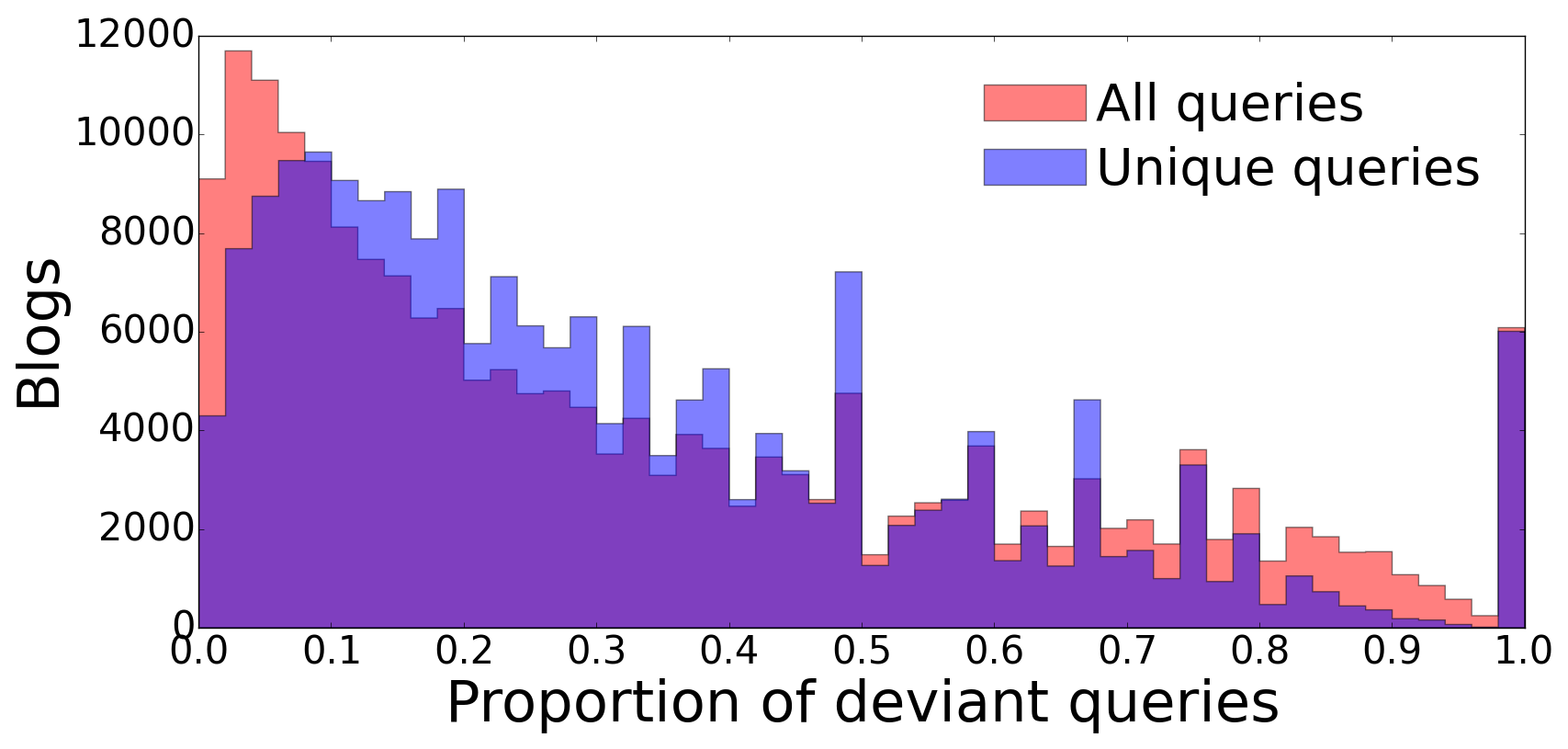}
\caption{Distribution of deviant query volume ratio reaching deviant nodes.}
\label{fig:pornitude}
\end{figure}

Figure~\ref{fig:query_expansion_iterations} shows that the {\em Deviant Graph Extraction} procedure converges quickly. We stop after 6 steps with $198K$ nodes hit by $4.2M$ unique queries. The final vocabulary containing $7,\!361$ words is made publicly available to the research community\footnote{\url{https://github.com/hpclab/DevCommunities/}}. In Figure~\ref{fig:pornitude} we report the distribution of the {\em deviant query volume} ratio for the deviant nodes detected. The distribution is skewed, showing that about 30\% of the nodes are hit by a majority of deviant queries.

To study the interaction of deviant nodes with the rest of the social network, we extracted a subset of the Tumblr follower and reblog networks with a snowball expansion starting from the $198K$ identified deviant nodes up to 3-hops away.
The follower is a snapshot of the graph done in December 2015; the reblog network was built from the reblog activity happened in the same month. Statistics about the resulting networks are reported in Table~\ref{tab:network_stats}.

We also obtained information about self-declared age and gender for about $1.7 M$ Tumblr users and, in particular, for about $10\%$ of the detected {\em deviant nodes}. The datasets include exclusively interactions between users who voluntarily opted-in for such studies.
All the analysis we report next has been performed in aggregate and on anonymized data.

 \section{Deviant graph in context} \label{sec:results}

\begin{table}[tp]
\scriptsize
\begin{tabular}{c|c@{\hskip 3mm}c@{\hskip 3mm}c@{\hskip 3mm}c@{\hskip 3mm}c@{\hskip 3mm}c@{\hskip 3mm}c@{\hskip 3mm}c}
& $|N|$ & $|E|$ & $\left\langle k\right\rangle$ & $D$ & $\rho$ & $C$ & $\overline{spl}$ &  $d$ \\
\hline
\textbf{All R} & 14M & 472M & 33 & $2 {\cdot} 10^{-6}$ & 0.06 & - & - & - \\
\textbf{All F} & 130M & 6,892M & 53 & $4 {\cdot} 10^{-7}$ & 0.10 & - &  - & - \\
\hline
\textbf{Deviant R} & 105K & 1.4M & 13 & $1 {\cdot } 10^{-4}$ & 0.04 & 0.10
 &  3.73 & 11  \\
\textbf{Deviant F} & 135K & 24.6M & 182 & $1 {\cdot} 10^{-3}$ & 0.07 & 0.13
 & 2.80 & 8 \\
\hline
\textbf{Prod$_1$ R} & 48K & 914K & 19 & $4 {\cdot} 10^{-4}$ & 0.04 & 0.09 & 3.44 & 9  \\
\textbf{Prod$_2$ R} & 16K & 305K & 19 & $1 {\cdot} 10^{-3}$ & 0.05 & 0.13 & 3.19 & 8  \\
\textbf{Bridge$_1$ R} & 9K & 36K & 4 & $5 {\cdot} 10^{-4}$ & 0.04 & 0.08 & 4.18 & 13  \\
\textbf{Bridge$_2$ R} & 3K & 32K & 11 & $4 {\cdot} 10^{-4}$ & 0.06 & 0.21 & 3.32 & 10  \\
\end{tabular}
\caption{Network statistics for the reblog (R) and follow (F) networks of the full graph sample (\textit{All}), the deviant graph (\textit{Deviant}), and the four communities that compose it (\textit{Producers$_{1,2}$} and \textit{Bridge$_{1,2}$}). All the statistics are about the giant weakly connected components and count only links whose both endpoints are in the considered node subset. $\left\langle k\right\rangle$=average degree, $D$=density, $\rho$=reciprocity, $C$=clustering, $\overline{spl}$=average shortest path length, $d$=diameter.}
\label{tab:network_stats}
\end{table}

The availability of data about the interaction between deviant nodes and the social network that surrounds them provides the unique opportunity to study the structure and dynamics of a deviant network within its context. We first analyze the shape of the deviant network and measure its connectivity with the rest of the social graph (Section~\ref{sec:results:community}). We then look into how the information originating from deviant networks spreads across the boundaries of the deviant group (Section~\ref{sec:results:diffusion}). Last, we study some demographic properties that characterize producers and consumers (Section~\ref{sec:results:demographics}).

\subsection{Deviant network connectivity} \label{sec:results:community}

The deviant network is a tiny portion of the whole graph, representing about 0.7\% of all the nodes in the reblog graph and 0.1\% of those in the follow network. So few nodes could be scattered along the social network or clustered together. So we ask:

\hbox{}
\noindent\textit{Q1) Are deviant nodes organized in a community?}

\noindent We consider the deviant networks as the subgraphs of the
follow and reblog Tumblr networks induced by the \textit{deviant nodes}. A
directional link in the follow (reblog) network from node $i$ to node
$j$ exists if $i$ follows (or reblogs the posts of) $j$, meaning that
the information flows from $j$ to $i$. Basic network statistics on such
subgraphs reveal that the deviant networks are quite dense, yet they have a high diameter (Table~\ref{tab:network_stats}).
Similar statistics have been observed before in other social
networks~\cite{aiello10link} and might be an indication of the
presence of strong sub-groups patterns, as well as a signal of the
absence of a community structure. To better determine the reason for such elongated shape, we run the Louvain
community detection algorithm~\cite{blondel2008fast} on the deviant
network\footnote{Louvain is a modularity-based graph clustering
algorithm that shows very good performance across several
benchmarks~\cite{fortunato10community} and that is fast to compute even
on large networks.}. Four clusters emerge, whose network statistics are
summarized in the bottom lines of Table~\ref{tab:network_stats}. To
determine their nature, we manually inspected the content of $250$ blogs
in each of them. More than $90\%$ of all the blogs in the two largest
clusters contain blogs that \textit{exclusively} produce explicit adult content,
aimed at an heterosexual public (\textit{Producers}$_1$) or at a male
homosexual public (\textit{Producers}$_2$). The blogs in the two
remaining communities post less explicit adult content and more sporadically, often by means of reblogging. They either focus on celebrities (\textit{Bridge}$_1$), or function as aggregator blogs with
high content variety, including depiction of nudity
(\textit{Bridge}$_2$).

\begin{figure}[tp]
\centering
\includegraphics[clip=true, width=.85\columnwidth]{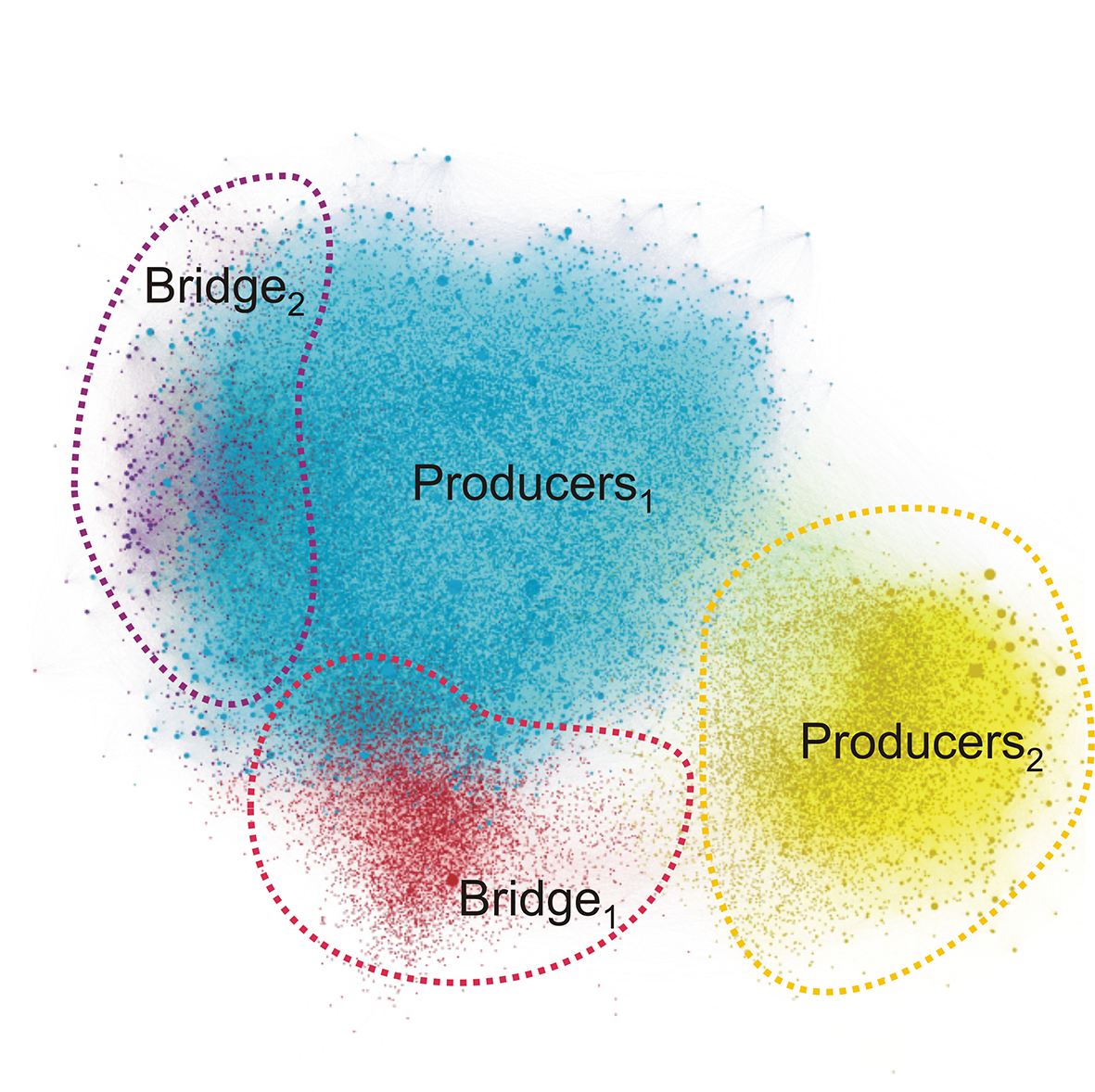}
\caption{Bird-eye view of the deviant network, with colors denoting algorithmically-extracted communities}
\label{fig:network_map}
\end{figure}

From a bidimensional visualization of the network layout
(Figure~\ref{fig:network_map}) it becomes apparent that the two bigger
clusters are two well-separated cores that give a characteristic hourglass shape to
the network, reason for the high diameter observed. The remaining
communities are peripheral and arranged in a crown-like fashion (which explains their high diameter) around
the largest sub-cluster \textit{Producers$_1$}. We name
the two smaller groups \textit{bridge communities} as their main focus
is not on deviant content but they are an entry point for deviant query
traffic and, as we shall see next, act also as bridges towards the rest
of the graph.

In short, we find that deviant nodes are not scattered in the social
network but are tightly organized in a structure of distinct
communities. To find out about the nature of their interaction with the
rest of the social ecosystem, we proceed to answer the next question.

\hbox{}
\noindent\textit{Q2) To what extent is the deviant graph connected to the rest of the social network?}

\noindent There are several ways to estimate the connectivity
between two sets of nodes in a graph. We use different metrics to
measure it between the four communities of the deviant
network and the rest of Tumblr, as summarized by the matrices in
Table~\ref{tab:matrices}; rows represent the group of nodes from which
the social tie originates, columns those on which it lands.

The average volume of connections (Table~\ref{tab:matrices}, left) provides a first indication about the
difference in connectivity across different groups. The diagonal has the
highest values because of the community structure of the deviant network
and of its sub-communities: members of a group have many more ties
towards other group members rather than to the outside. This is true in
particular for the two \textit{Producer} clusters. The volume of links
incoming to the largest producer cluster is particularly high from the
smallest bridge community ($Bridge_2$), which surrounds it. The average
Tumblr user in our sample follows around $51$ users, between $2$ or $3$ of which
are in the core of the deviant network and around $2$ of them are in
bridge communities; similarly, among the $33$ users reblogged in one month
by the average user, one is from a \textit{Producer} cluster and one from a \textit{Bridge} group.

When looking at raw volumes, the amount of links from the deviant
network to the rest of the graph is very high, mainly due to the high
dimensionality of the set of nodes that are not deviant. To partially
account for dimensionality of the groups, we measure the connectivity
with density computed as the ratio of edges between the two groups over
the total number of possible edges between them (Table~\ref{tab:matrices}, center).
Also in this case the overall patterns hold, but the connectivity towards the external graph
drops significantly.

Values of density are still affected by size, though. It is known that
in real networks there is a strong correlation between density and number of nodes~\cite{leskovec05graphs}.
To fix that, in the spirit of established work
in complex systems~\cite{schifanella10folks} we resort to a comparison of the real
network connectivity with a \textit{null model} that randomly rewires the links
while keeping the degree of each node unchanged. The values we report in
Table~\ref{tab:matrices} (right) indicate how many times the number of
connections observed deviate from the null model. Also in this case,
values on the diagonal are very high (except for the outer network,
which has a value close to 1, as expected). Also, this computation
highlights that ordinary users have a tendency to reblog content from the core of
the deviant network almost 7 times more than random and between 16 and
53 times more than random from the bridge community members.

In summary, the core of the deviant community is dense but it is far
from being separated from the rest of the graph, which is connected to
it both directly and even more tightly through bridge groups.

\begin{table*}[t!]
\centering
\label{netstats}
\begin{tabular}{cccc}
& Average volume & Density ($\cdot 10^{-2}$) & Null model comparison \\
\rotatebox{90}{\hspace{1.5cm}Follow} & \includegraphics[clip=true, width=.55\columnwidth]{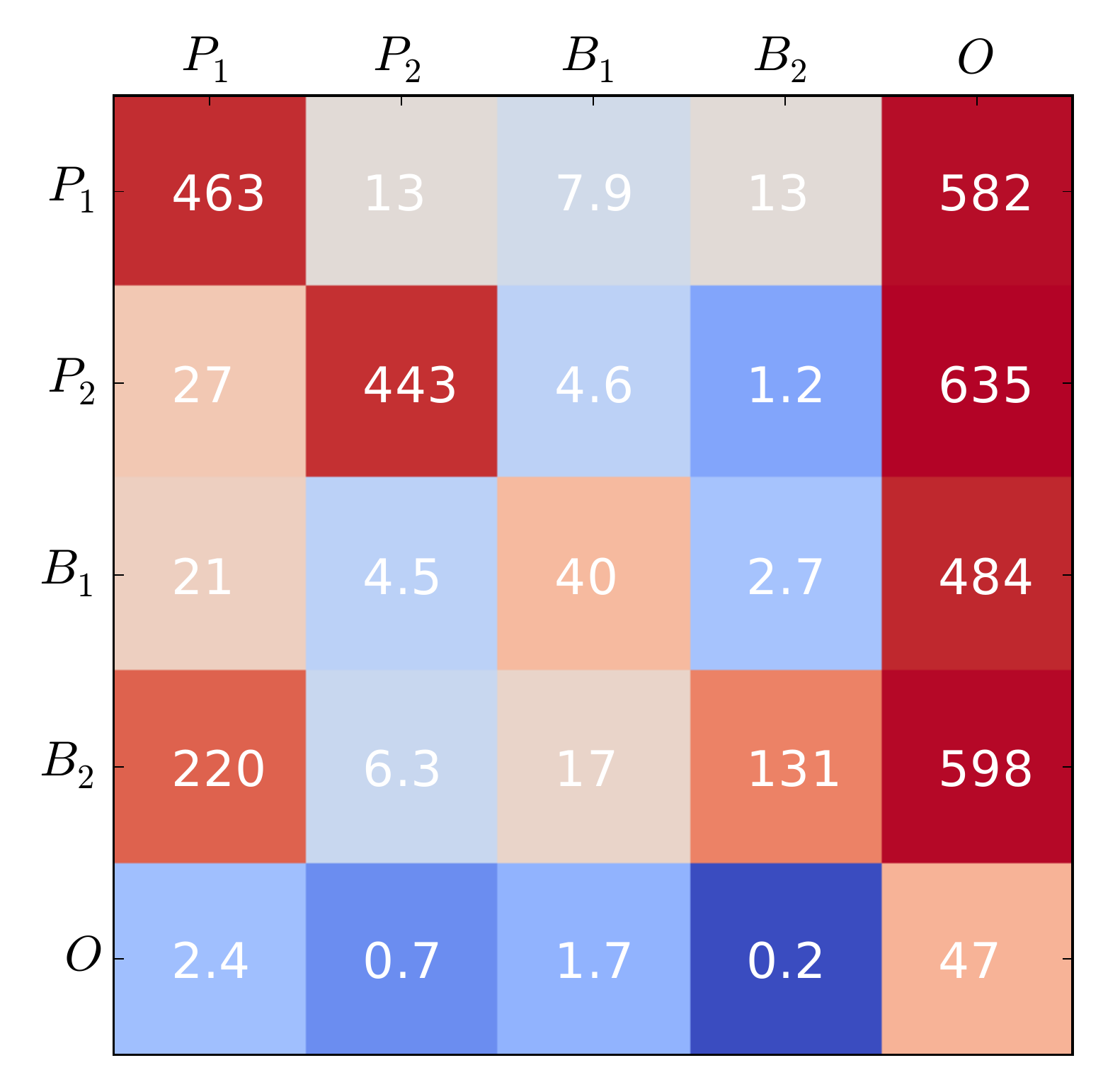} & \includegraphics[clip=true, width=.55\columnwidth]{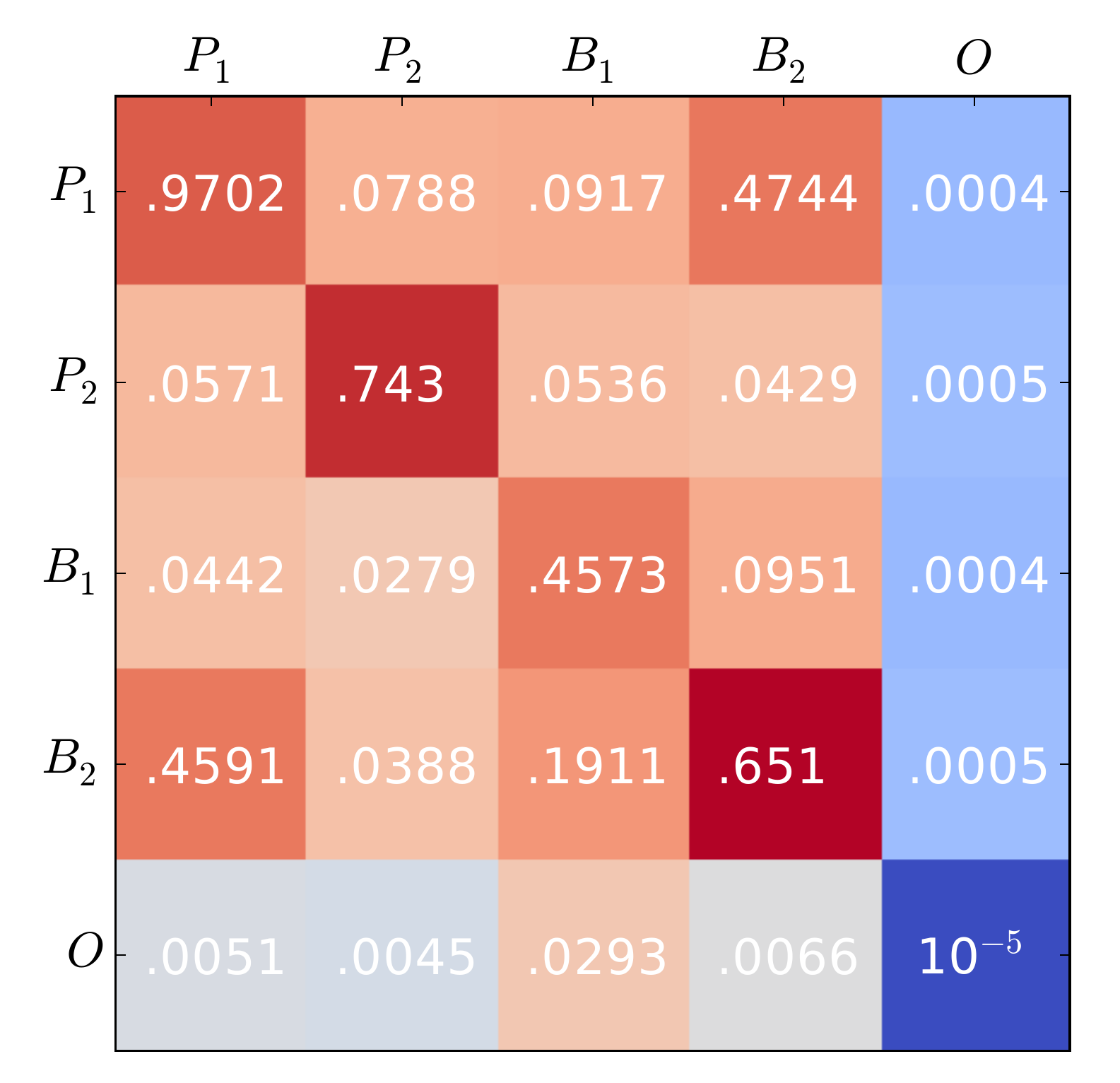} & \includegraphics[clip=true, width=.55\columnwidth]{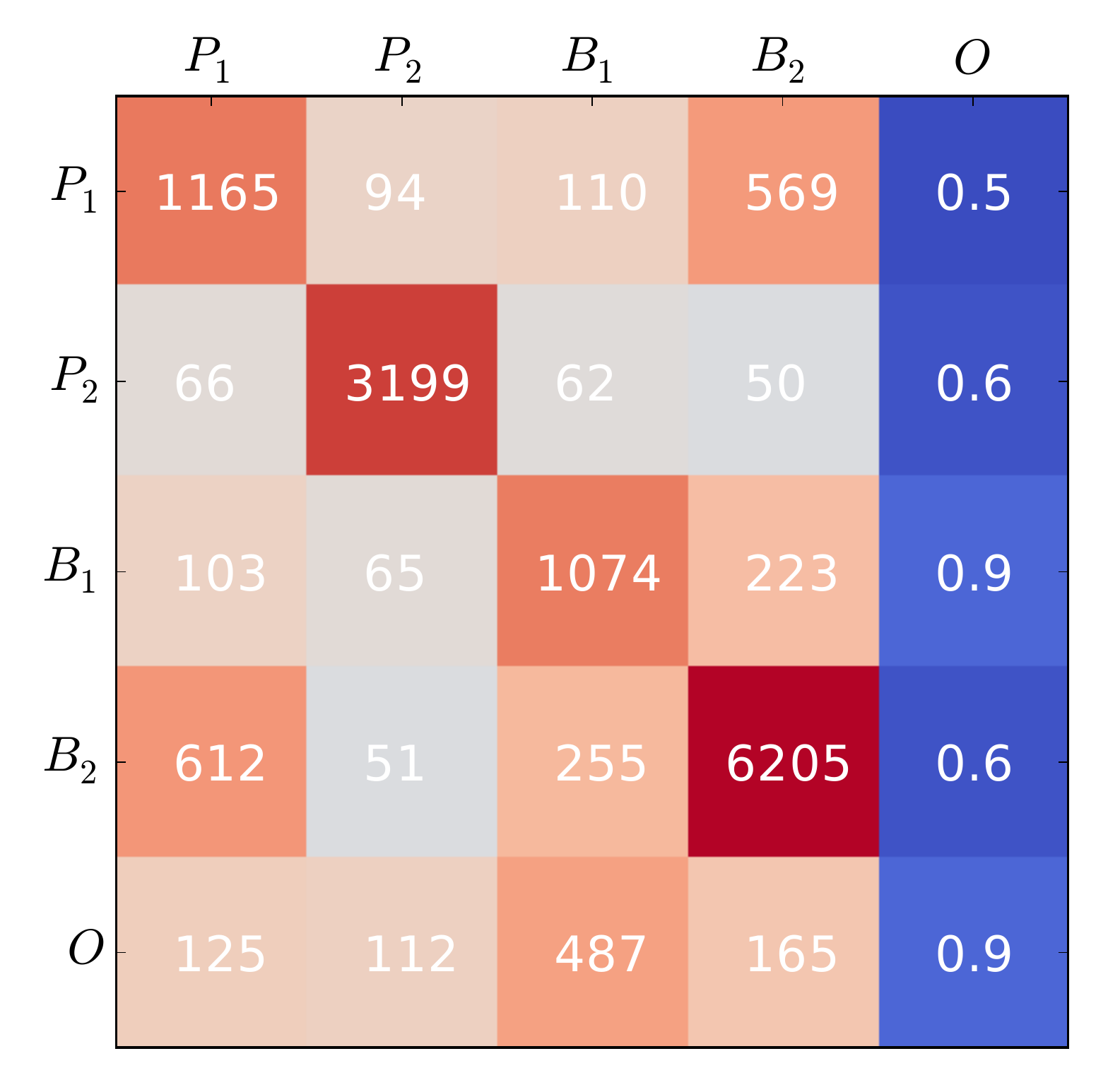}\\
\rotatebox{90}{\hspace{1.5cm}Reblog}& \includegraphics[clip=true, width=.55\columnwidth]{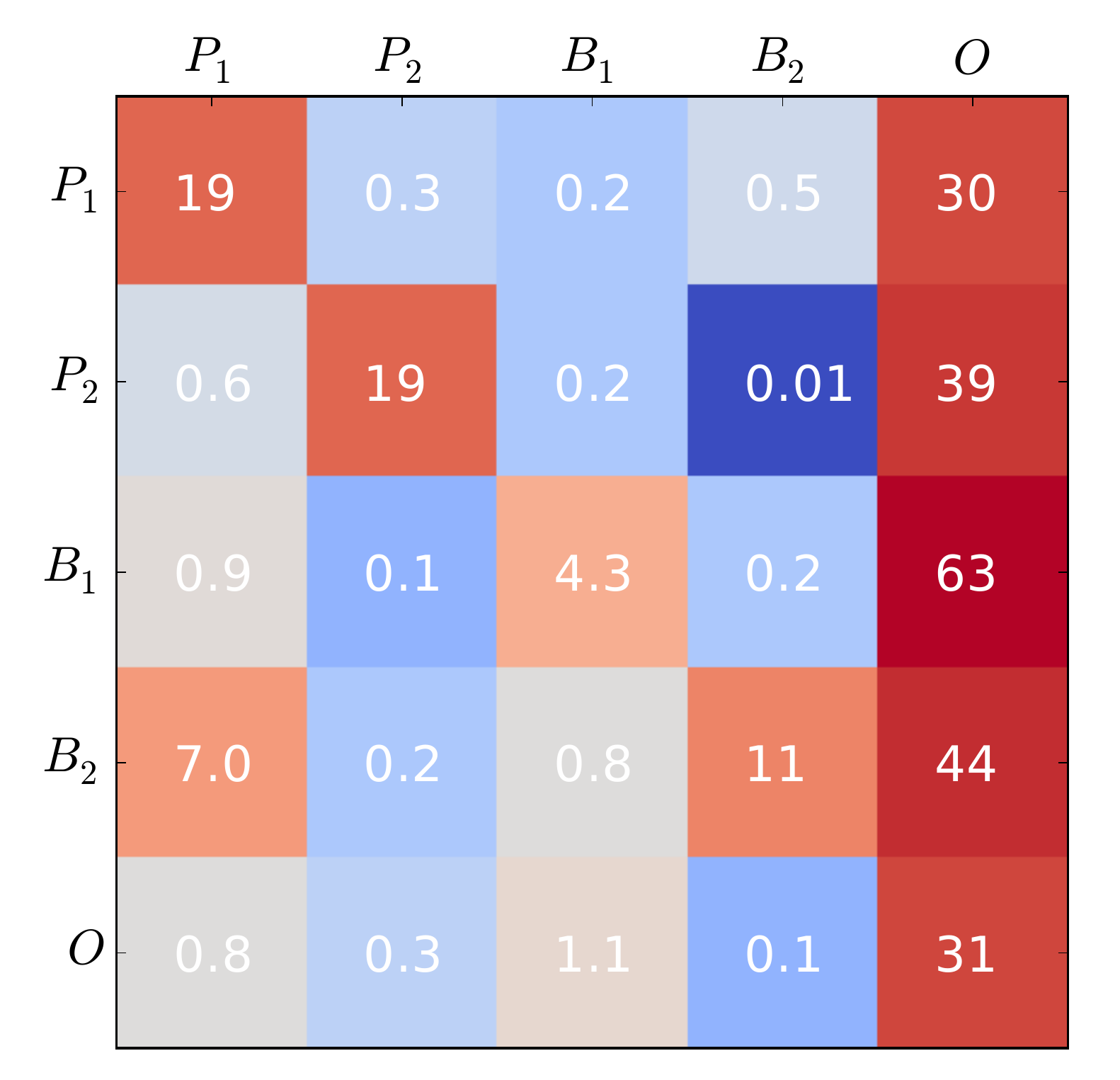} & \includegraphics[clip=true, width=.55\columnwidth]{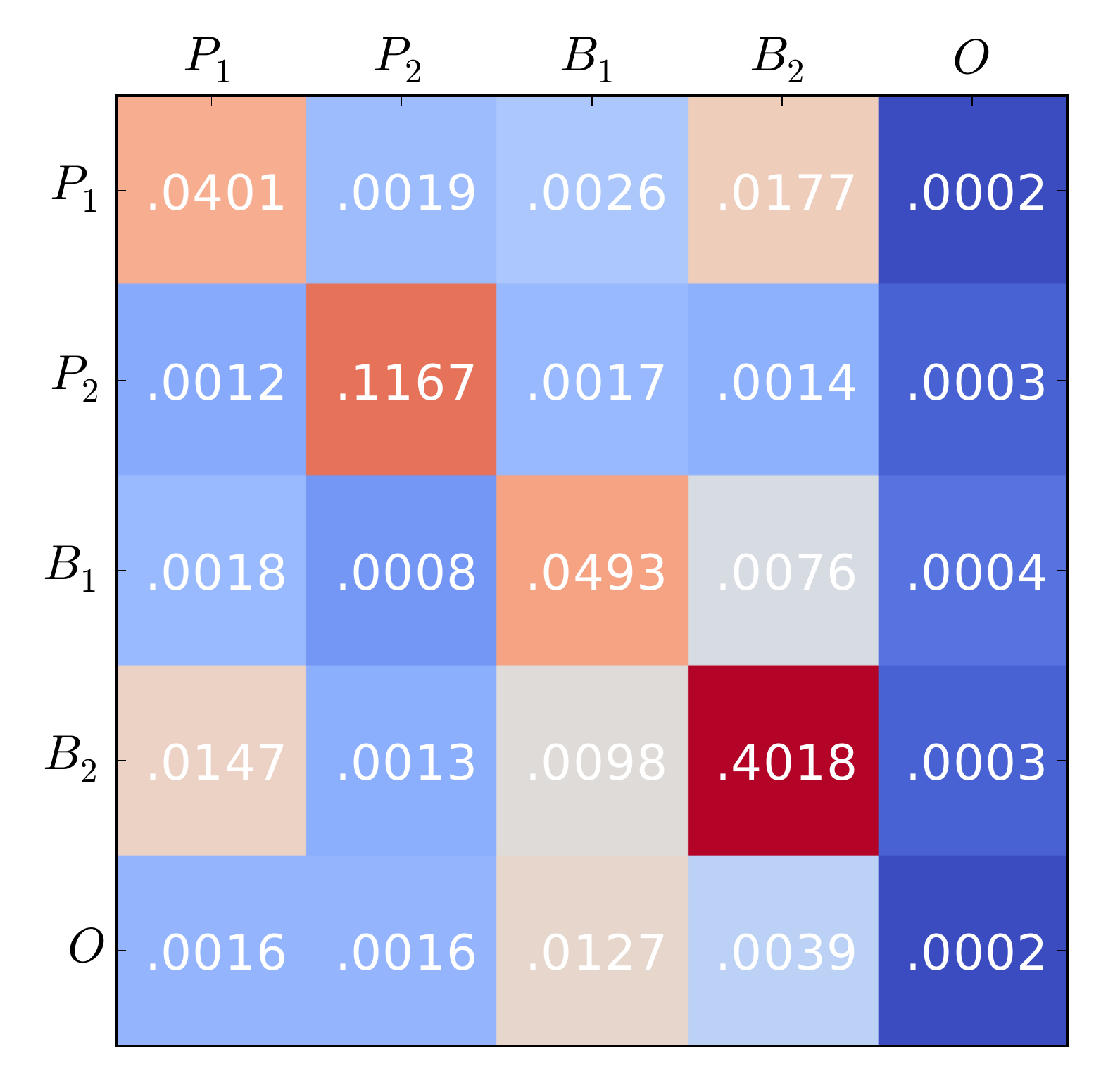} & \includegraphics[clip=true, width=.55\columnwidth]{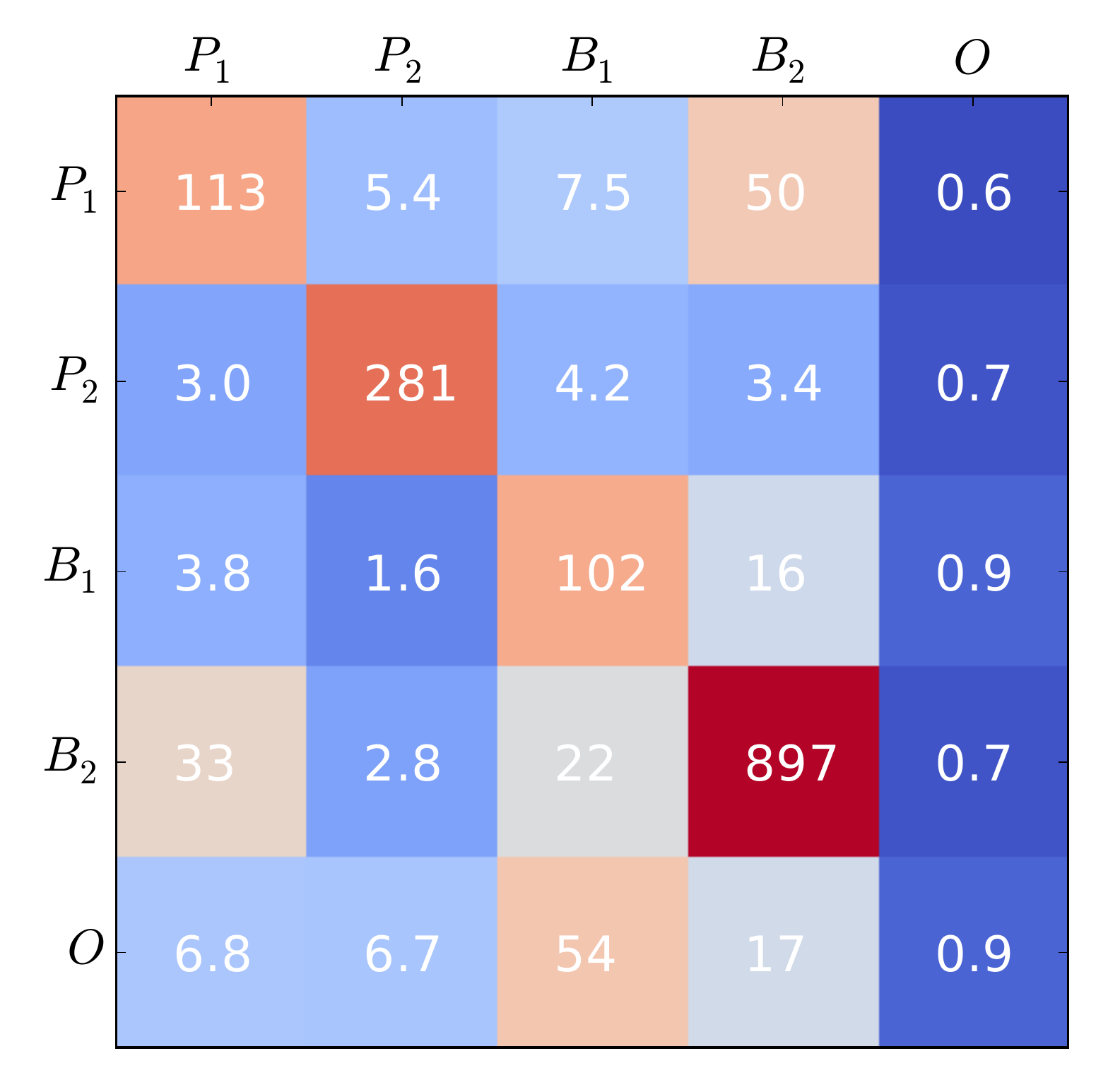}\\
\end{tabular}
\caption{Measures of connectivity between the communities in the deviant network (\textit{Producers} $P_1$, $P_2$ and \textit{Bridges} $B_1$, $B_2$) and the rest of the social network $O$, for both the follow (top) and reblog (bottom) relations. Link directionality is considered: ties originate from groups listed on the rows and land on groups listed on the columns.}
\label{tab:matrices}
\end{table*}

 \subsection{Deviant content reach} \label{sec:results:diffusion}

We found that, although the deviant network forms a tightly connected community, it is not isolated from the rest of the social graph. This calls for an investigation about the visibility that the deviant content has in the outer network and what are the main factors that determine its exposure. We do so by answering the three research questions below.

\hbox{}
\noindent \textit{Q3) How much deviant content spreads in the social graph and who are the main agents of diffusion?}

The exposure to deviant content goes beyond the members of the deviant network who are the \textit{producers} of original adult material. Specifically, the \textit{consumers} of deviant content can be categorized in three classes. The first is the class of \textit{active consumers}: nodes who reblog (but not necessarily follow) adult posts, thus contributing to its spreading along social ties. Posts can be re-blogged in chains and create diffusion trees that potentially spread many hops away from the original content producer, therefore active consumers could further be partitioned in those who spread the content \textit{directly} from the producers and those who do it with \textit{indirect} reposts. The second is the class of \textit{passive consumers}: nodes who do not contribute to the information diffusion process but are explicitly interested in adult content because they directly follow the producer nodes. The last class is the one of \textit{involuntary consumers} (or \textit{unintentionally exposed} users): users who do not follow any producer node and do not reblog their content, but happen to follow at least one active consumer who pushes adult content in their feed through reblogging.

By drawing a quantitative description of the volume of deviant content reaching these three classes we can estimate how much the adult community is visible in the network at large. We adopt a conservative approach in which we consider the two \textit{Producers} communities as the only ones generating original explicit (homosexual and heterosexual) content. Given the results of the aforementioned manual inspection, we are very confident that their activity is completely focused on the production of adult material.

\begin{figure}[tp]
\centering
\includegraphics[clip=true, width=.85\columnwidth]{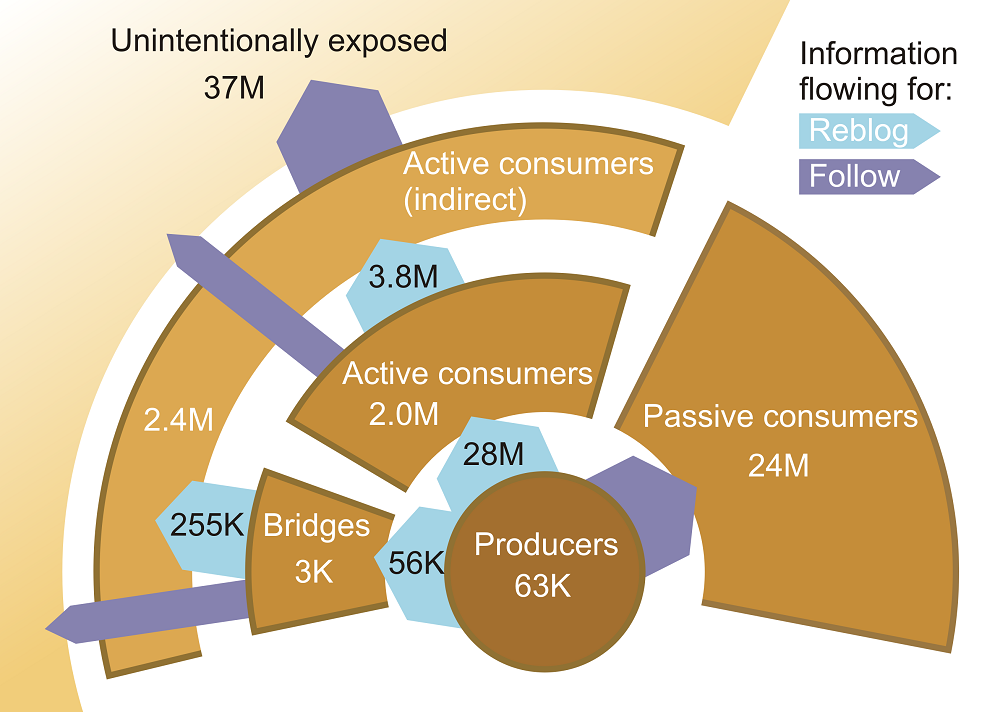}
\caption{Diffusion of deviant content from the core of Producers to the rest of the network. Sectors represent disjoint user classes and arrows encode the information flow between them. Reblog arrows report the total volume of reblogs between two classes.}
\label{fig:diffusion_map}
\end{figure}

We measure the size of the different consumer classes and the amount of content that flows through or to them by means of reblogging. The results are summarized by the schema in Figure~\ref{fig:diffusion_map}. The network of deviant content producers is very small but receives a considerable amount of attention from direct observers. The audience of passive consumers counts almost 24M people. Around 2M users reblog directly from the deviant network, for a total of around 28M reblog actions in one month. A consistent part of the two \textit{Bridge} communities within the deviant graph (a total of 3K users) are also direct consumers, and they reblog \textit{Producers} 56K times per month. When looking at the set of 2.4M users who indirectly reblog deviant content, we see that only a small fraction of their monthly reblogs (less than 7\%) is performed through bridge communities. However, in relative terms, bridge communities are considerably more efficient in spreading information than the average active consumer. If we consider efficiency $\eta$ of a user set $U$ as the ratio between reblogs done $r_d$ and reblogs received $r_r$, weighted by the cardinality of the set $\left( \eta=\frac{r_r}{r_d \cdot |U|} \right)$, we discover that the bridge communities ($\eta=1.5\cdot10^{-3}$) are several orders of magnitude more effective in spreading the content farther away in the network than the rest of active consumers ($\eta=6.7\cdot10^{-8}$). Last, the audience of users who are potentially exposed in an unintentional way to deviant content includes almost 40M people. This figure should be considered as an upper bound on the number of people who actually have been exposed, as a follower of an active consumer might not see the pieces of deviant content for a number of reasons (e.g., inactivity, amount of content in the feed). That said, the pool of people who are potentially exposed is still very wide.

\hbox{}
\noindent \textit{Q4) What is the perception of deviant content consumption from the perspective of individual nodes?}

Similar to real life, individuals in online social networks are most often aware of the activities of their direct social connections only but lack a global knowledge of the behavior of the rest of the population. In fact, the broad degree distribution of social networks may lead to the over-representation of rather rare nodal features when they observed in the local context of an ego-network. This phenomenon has been observed in the form of the so-called \textit{friendship paradox}~\cite{feld91your,hodas13friendship}, a statistical property of social networks for which on average people have fewer friends than their own friends. More recently the concept has been extended by the so-called \textit{majority illusion}~\cite{lerman15majority}, which states that in a social network with binary node attributes there might be a systematic local perception that the majority of people (50\% or more) possess that attribute even when it is globally rare. As an illustrative example, in a network where people drinking alcohol are a small minority, the local perception of most nodes can be that the majority of people are drinkers just because drinkers happen to be connected with many more neighbors than the average. In our case study, active deviant content consumption is definitely a minority behavior compared to the 130M users in our sample.

\begin{figure}[tp]
\centering
\includegraphics[clip=true, width=.99\columnwidth]{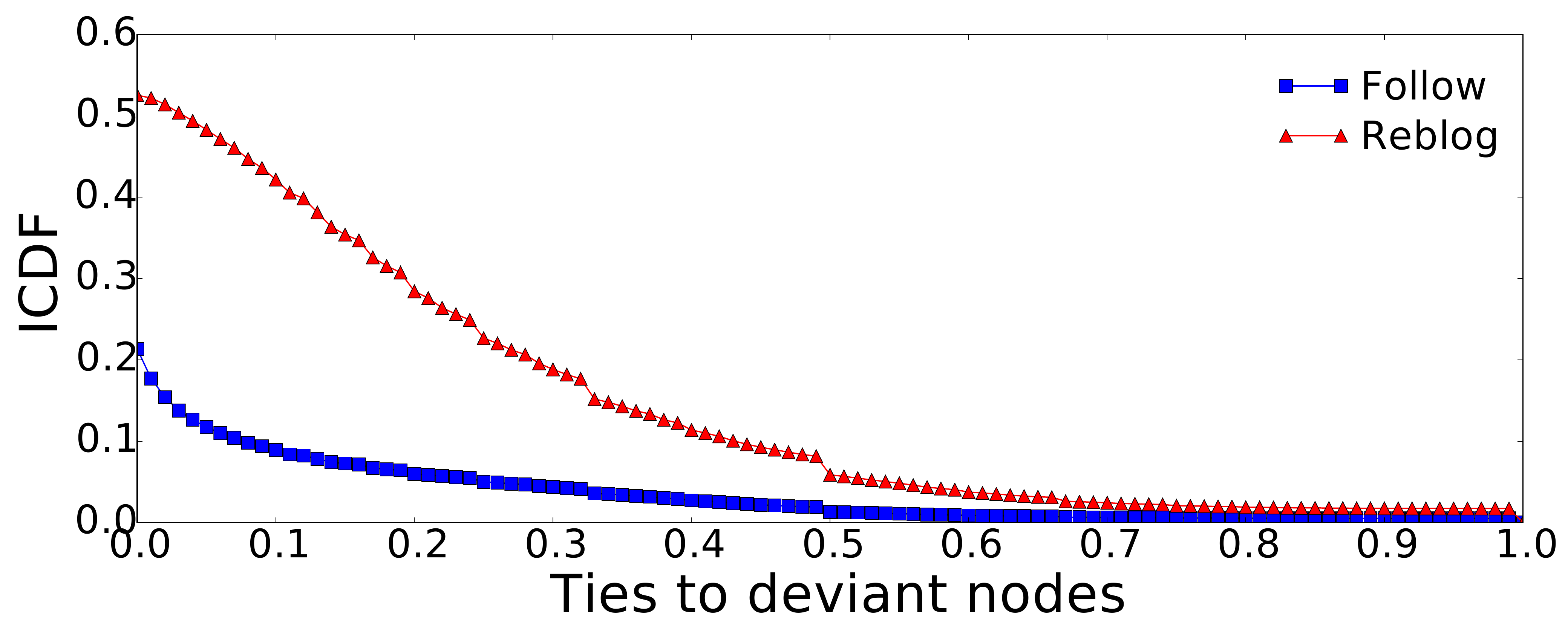}
\caption{Proportion of nodes with at least a given ratio of outlinks landing on deviant nodes (inverse cumulative density function).}
\label{fig:majority_illusion}
\end{figure}

To estimate the presence of any skew in the local perception of deviant content consumption, we consider the nodes who are not producers and calculate the distribution of the proportion of their neighbors (in both the follow and reblog graphs) that either produce or reblog deviant material. The result is summarized in Figure~\ref{fig:majority_illusion}. We observe that the follower network is nowhere close to exhibit the majority illusion phenomenon, with only the $10\%$ of the population having $10\%$ or more of their neighbors posting or reblogging deviant content. The effect increases sensibly when considering the reblog network, with $40\%$ of the population locally observing more than $10\%$ of their contacts reblogging deviant content and almost $10\%$ having more than half of their neighbors doing it. This happens partly because the size of the reblog network is one order of magnitude smaller than the one of the follower network, as we consider reblogging activity for one month only. Still, this means that when looking at recent activity only, local perception biases are much stronger (although not predominant) in the community than what can be inferred from the static follow graph.

Although strongly biased perceptions are not predominant when counting the number of neighbors, a stronger bias emerges when looking at the \textit{volume} of deviant content that is observed by a node from its neighbors. More than 71\% of nodes reblogs less deviant content than the average of their friends (considering friends who posted or reblogged at least once in the time frame we consider). This effect, that derives directly from the strong correlation between degree and number of posts and reblogs, suggests that the local users' perception of other people's behavior is skewed towards an image of pervasive consumption of deviant content.

\hbox{}
\noindent \textit{Q5) Is it possible to reduce the diffusion of deviant content with targeted interventions?}

Previous literature that investigated the properties of small-world networks indicates that information spreading or other phenomena of contagious nature can be drastically reduced by acting on a limited number of nodes in the graph~\cite{pastor05epidemics}. Effectiveness of targeted interventions has been shown in a variety of domains, epidemics being the most prominent among them.

The intuition informed by previous work suggests that the wide diffusion of deviant content can be reduced by properly marking the posts produced by a small set of core nodes and showing them only to people who explicitly declared their interest for that specific topic. In a simplified experimental scenario, we measure the proportion of active consumers reached by adult content in a setting where all the posts from a set of core nodes $C$ are erased. The question is how to select $C$ and how big it needs to be to uproot the diffusion process.

\begin{figure}[tp]
\centering
\includegraphics[clip=true, width=.99\columnwidth]{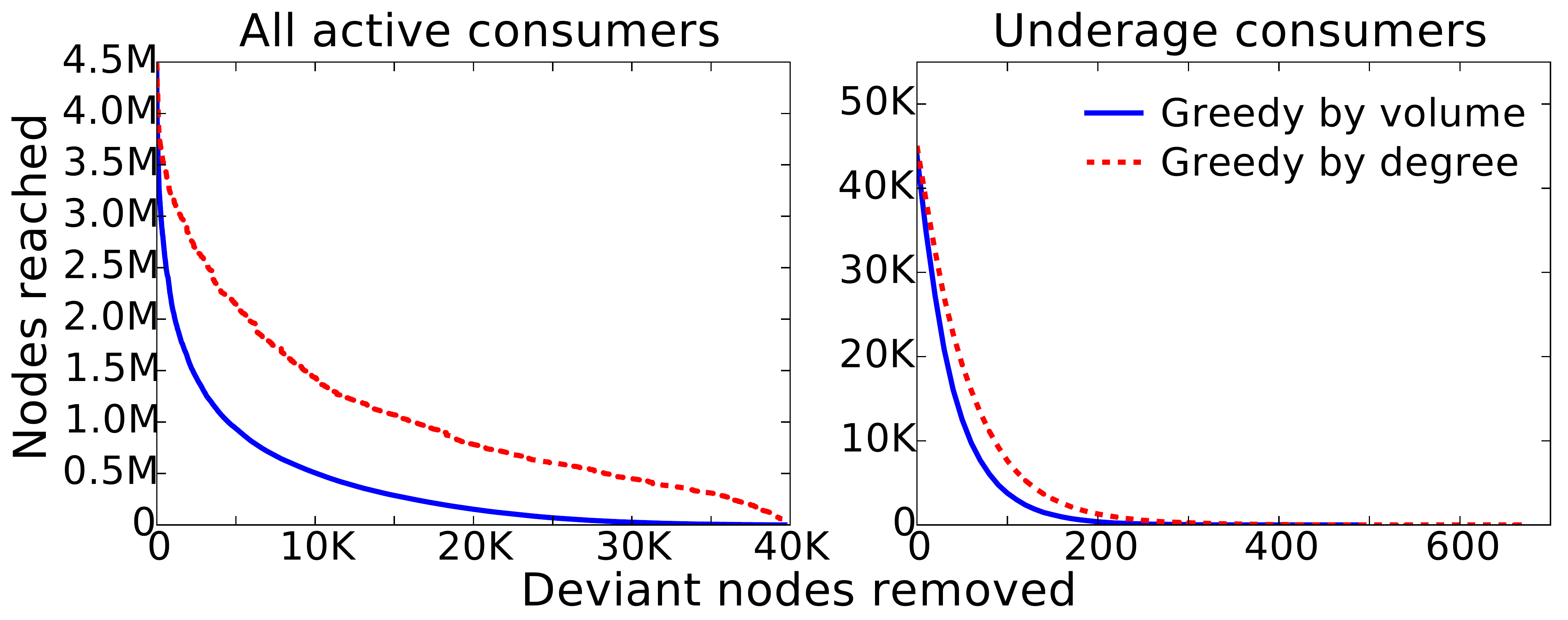}
\caption{Shrinkage of content diffusion after deviant nodes removal, using two different strategies.}
\label{fig:diffusion_prevention}
\end{figure}

The optimal selection of nodes is a set cover problem (NP-complete), but we test two common approximated strategies to solve it: \textit{i)} \textit{greedy by volume}, an algorithm that ranks nodes by the number of blogs that are reached by the content they produce; and \textit{ii)} \textit{greedy by degree}, that takes into account the network structure only and ranks nodes by their in-degree in the reblog network. The effectiveness of the two approaches as $|C|$ increases in shown in Figure~\ref{fig:diffusion_prevention}. Although using the indegree as proxy for the diffusion potential is not optimal, the removal of the $5,\!000$ highest indegree nodes curbs the diffusion by more than $50\%$. As expected, the strategy by volume is more effective (as it better approximates the optimal set cover), with a surprisingly sharp decay of the deviant content reach. The removal of the $5,\!000$ top nodes reduces the information spreading by nearly $80\%$, which increases to almost $100\%$ when extending the block to $25,\!000$ nodes. Furthermore, using our sample of demographic information, we find that to limit the exposure of underage users would be sufficient to remove the $200$ top nodes, as identified by any of the two selection strategies.

 \subsection{Demographics factors} \label{sec:results:demographics}

The demographic composition of online adult content consumers has been
measured by several sociological surveys (see
Section~\ref{sec:related}), but none of them partitions the
participants according to their type of consumption. Yet, we have shown
that the categories of people exposed to online deviant content range
from the active content producers to unintentional consumers. This
calls for an investigation of the relationship between type of
consumption and demographic characterization.

\hbox{}
\noindent\textit{Q6) Is there a significant difference in the
distribution of age and gender between members of the deviant network
and people with different levels of exposure to deviant content?}

\begin{figure}[tp]
\centering
\includegraphics[clip=true, width=.99\columnwidth]{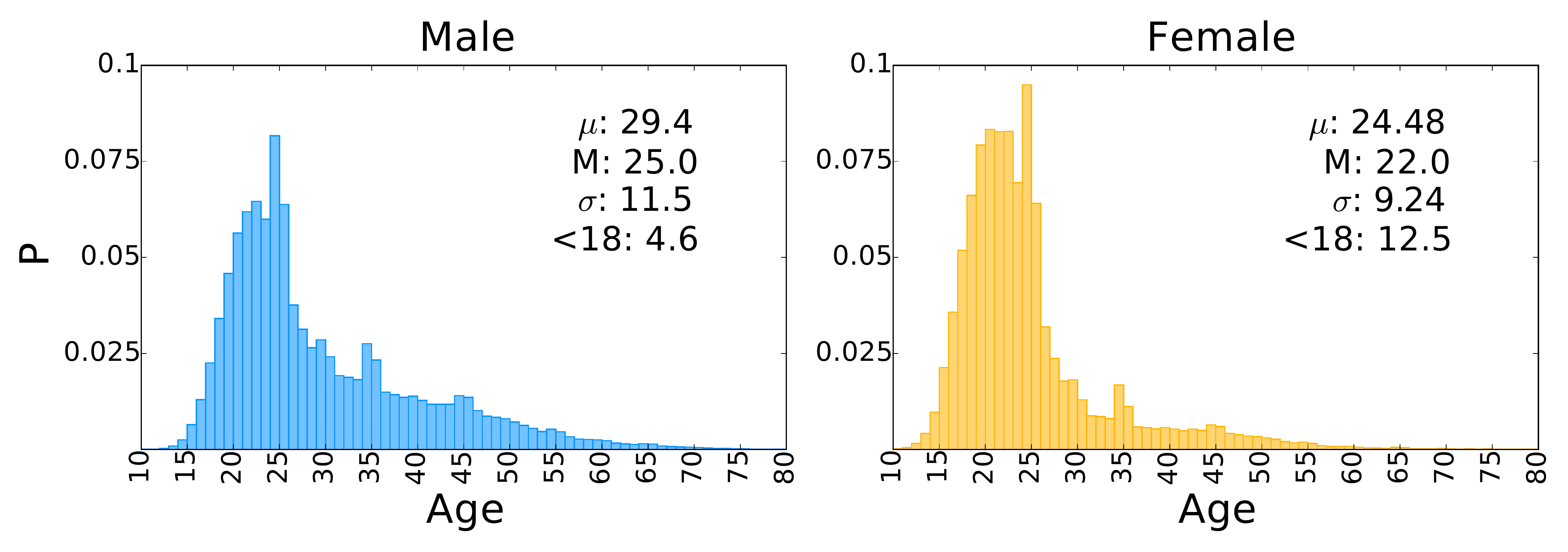}
\caption{Age distribution of Tumblr users in our dataset. Mean $\mu$, median M, standard deviation $\sigma$, and percentage of users under 18 years old are reported.}
\label{fig:age_total}
\end{figure}

\noindent We report the distribution of age and gender of users with different
levels of exposure to adult content, computed on the sample of 1.7M users who 
self-reported their demographic information. The average age in the sample
is slightly higher than 26, and female are the majority ($72\%$). To
partly validate the user-provided information, we first compare them
with third-party statistics. Our numbers are roughly compliant with
several public reports that rely on orthogonal methods for assessing the age and gender
of users (e.g., surveys and clickstream
monitoring~\cite{pingdom12report,kantar14social}). Those show that the
Tumblr user base is the youngest among the most popular social networks
and composed of women (65\%)~\cite{taylor2012women}. Also, we further validate the gender
data by assessing that the $95\%$ of users in the \textit{Producer$_2$} cluster
focused on male homosexual content are indeed male. The overall age
distribution of age by gender is shown in Figure~\ref{fig:age_total}:
male tend to be older, originating a distribution with a fatter tail between age 35 and 55.
Despite the spikes corresponding to birthdays in round decades (1970,
1980, and 1990), probably due to misreporting, the distribution still
tends to be Gaussian, as expected.

\begin{figure*}[tp]
\centering
\includegraphics[clip=true, width=.99\textwidth]{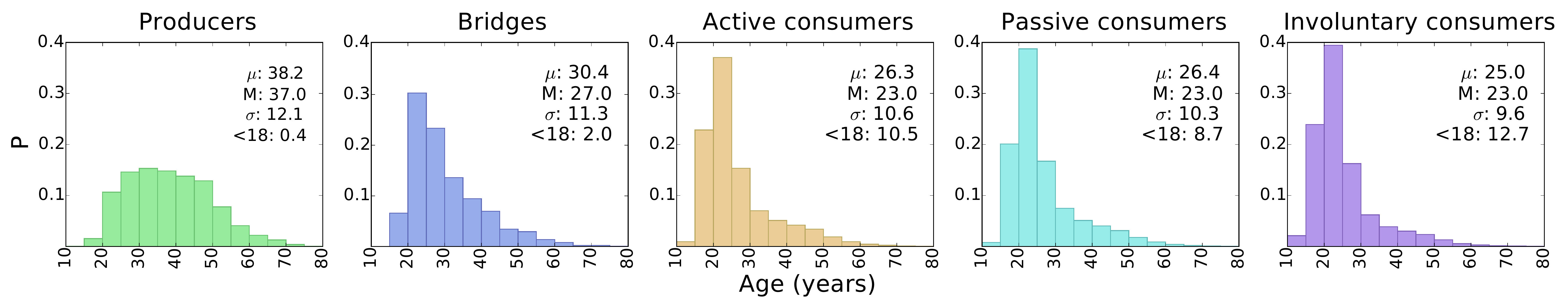}
\caption{Age distribution of different groups of producers and consumers of adult content.}
\label{fig:age_group}
\end{figure*}

We then measure differences in age\footnote{The number of samples in each age distribution is high; therefore, as expected, all the differences between the average values are statistically significant ($p<0.01$) under the Mann-Whitney test.} and gender distribution for the user classes of \textit{producers}, \textit{bridges}, \textit{active
consumers}, \textit{passive consumers}, and \textit{unintentionally
exposed} users (Figure~\ref{fig:age_group}). Producers are considerably
older than the typical user, averaging around age 38 and with almost
no underage users. Different from the overall distribution, they are 
mostly male ($82\%$), in alignment with studies indicating that
men are more involved in assiduous consumption of adult material. Bridge
groups are fairly gender-balanced (with more female --68\%--
in the celebrity-oriented community) and include younger people (30 years old on average).

Consumers of deviant nodes who actively reblog or passively follow
deviant blogs are covered by demographic data at $12\%$, proportion
that drops to $4\%$ among those who follow deviant nodes.
In both classes, the
age is quite representative of the overall Tumblr population in our
sample (about 68\% female). The same male-female proportion holds for
people that are potentially exposed to deviant content in an
unintentional way. This last class has the highest proportion of underage people
($13\%$), which reinforces the concern about young teens unwillingly seeing inappropriate content.

The fact that the gender distribution for active and passive consumers
deviates only slightly from the overall gender distribution is in partial disagreement with previous studies on gender and
sexual behaviour~\cite{hald06gender,kvalem2014self} which state that men
are usually more exposed than women to adult material. 

We conjecture that this might happen because of the tendency of female
to have their peak of adult content consumption in a much younger age
than men (as shown by~\cite{ferree2003women}), combined with the predominance of young
female among Tumblr users. To verify it, we aim to answer one last
question.

\hbox{}
\noindent \textit{Q7) Does age have an effect on how different genders
consume adult content?}

\noindent To find out, we measure the proportion of male and
female actively exposed to deviant content (by reblogging), by age.
We apply a min-max normalization to the obtained values so that scores towards $0$ ($1$) represent the minimum (maximum) level of engagement.
The curve for men shows an increasing trend that plateaus at its maximum in
the range of age 35 to 55. In contrast, women, although less exposed
than men at any age, have their peak in their 20s, much earlier than
men. This observation supports previous findings~\cite{ferree2003women}
and explains the distributions we observed.

\begin{figure}[tp]
\centering
\includegraphics[clip=true, width=.85\columnwidth]{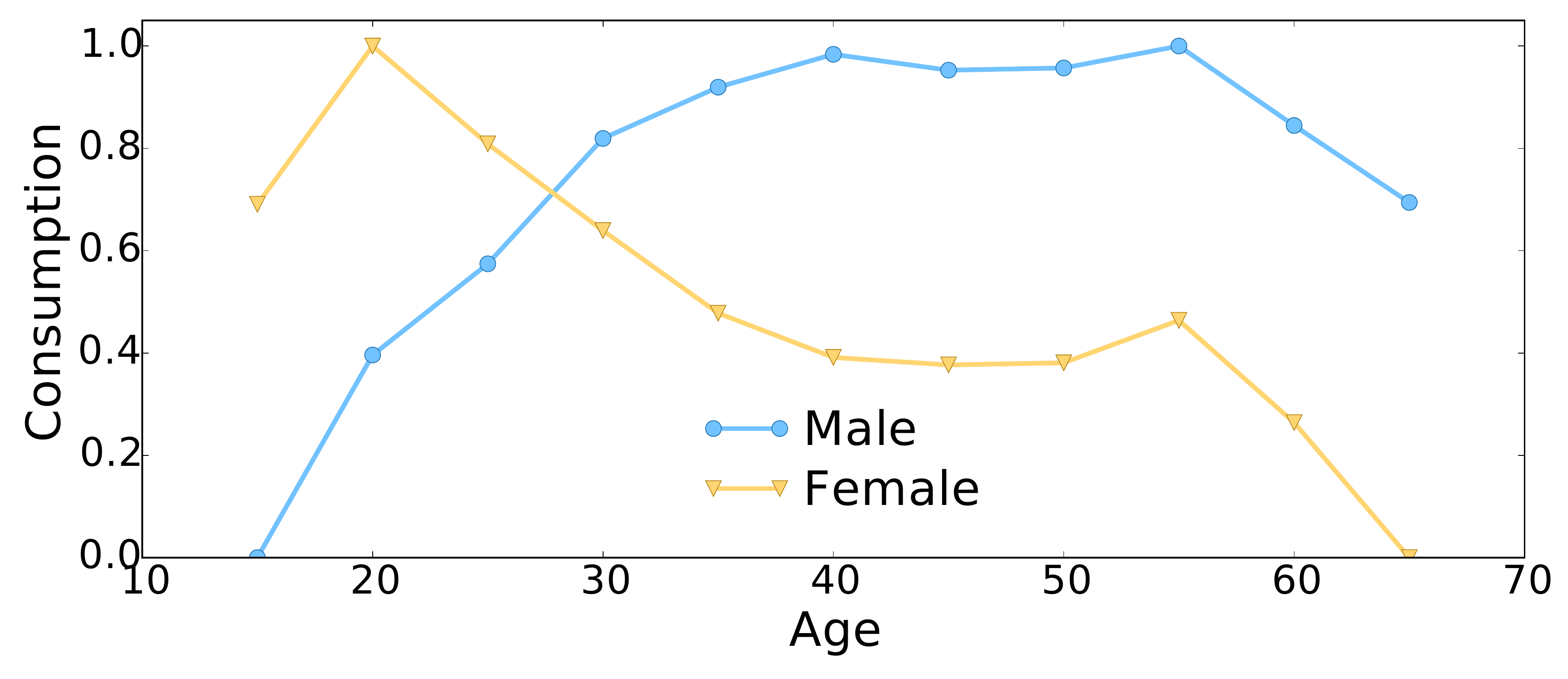}
\caption{Ratio of male and female consuming adult content for different age bands (min-max normalized).}
\label{fig:age_ratio}
\end{figure}

 \section{Conclusions} \label{sec:conclusion}

This work aims to motivate researchers who study all types of deviant communities online as well as offline to explore in more depth the interaction between the agents in such networks and the external social environment. Our contribution scratches only the surface of the exploration space that underlies the many types of deviant networks and the multitude of settings they are situated within. The study we have presented is limited under many aspects, beginning from the focus on a single type of deviant behavior --adult material consumption-- that is much more pervasive than others (e.g., anorexia) and, in that, has unique characteristics that likely cannot generalize to other deviant groups. In terms of methodology, alternative techniques (e.g., computer vision) could be used to identify adult content without a dedicated dictionary; those could possibly lead to describe the same phenomenon from a slightly different angle, for instance considering more exhaustively nodes that are not reached by search traffic. To address some of these points we plan to expand our study in both breadth and depth. In future work we will consider multiple online platforms (Twitter and Flickr being two ideal candidates, as they do not apply strong restrictions on the uploaded content) and multiple deviant network types at different scales (e.g., content advocating violent behavior within the adult community). Also, we plan on analyzing the temporal dynamics of the deviant content spreading along social links.

Yet, we believe that our study has already important theoretical implications in revealing, for the first time on very large scale, that deviant communities can be deeply rooted into the relational fabric of a social network, and that the echo of their abnormal activity can reach a plenitude of ordinary users. Also, from a practical point of view, learning the effect that a minority group can have on a much larger audience is key to trigger mechanisms able to contain risky deviant phenomena by means of targeted interventions on few nodes, as we have shown. We believe that this work could set the basis for a line of study that could lead to a deeper understanding of deviant networks and of their impact on everyone's life.
\section*{Acknowledgments}
\small{This work was partially supported by the EC H2020 Program INFRAIA-1-2014-2015 {\em SoBigData: Social Mining \& Big Data Ecosystem} (654024).}
\balance
\bibliographystyle{aaai}
\bibliography{bibliography}

\end{document}